\documentclass[pra,twocolumn,superscriptaddress]{revtex4-1}
\pdfoutput=1
\usepackage{amsfonts,amssymb,amsmath,bbm}
\usepackage[utf8]{inputenc}
\usepackage[english]{babel}
\usepackage[T1]{fontenc}
\usepackage{amsmath}
\usepackage{hyperref}
\usepackage{graphicx}
\usepackage{subcaption}
\usepackage{tikz}
\usepackage{lipsum}
\usepackage{mathpazo} 
\usepackage{amsthm}
\usepackage{epsf}
\usepackage{float}

\newtheorem{definition}{Definition}
\newtheorem{coro}{Corollary}

\newcommand{\be}{\begin{equation}}
\newcommand{\ee}{\end{equation}}
\newcommand{\beq}{\begin{eqnarray}}
\newcommand{\eeq}{\end{eqnarray}}

\newtheorem{prop}{Proposition}
\newtheorem{obs}{Observation}

\newcommand{\tr}{\text{Tr}}
\newcommand{\ket}[1]{|#1\rangle}

\def\squareforqed{\hbox{\rlap{$\sqcap$}$\sqcup$}}
\def\qed{\ifmmode\squareforqed\else{\unskip\nobreak\hfil
\penalty50\hskip1em\null\nobreak\hfil\squareforqed
\parfillskip=0pt\finalhyphendemerits=0\endgraf}\fi}
\def\endenv{\ifmmode\;\else{\unskip\nobreak\hfil
\penalty50\hskip1em\null\nobreak\hfil\;
\parfillskip=0pt\finalhyphendemerits=0\endgraf}\fi}

\begin{document}

\title{State independent contextuality advances one-way communication}


\author{Debashis Saha}
\email{saha@cft.edu.pl}
\affiliation{Institute of Theoretical Physics and Astrophysics, National Quantum Information Centre, Faculty of Mathematics, Physics and Informatics, University of Gda\'{n}sk, 80-952 Gda\'{n}sk, Poland}
\affiliation{Center for Theoretical Physics, Polish Academy of Sciences, Al. Lotnik\'{o}w 32/46, 02-668 Warsaw, Poland}
\author{Pawe\l{} Horodecki}
\affiliation{Faculty of Applied Physics and Mathematics, National Quantum Information Center, Gda\'{n}sk University of Technology, 80-233 Gda\'{n}sk, Poland}
\author{Marcin Paw\l{}owski}
\affiliation{Institute of Theoretical Physics and Astrophysics, National Quantum Information Centre, Faculty of Mathematics, Physics and Informatics, University of Gda\'{n}sk, 80-952 Gda\'{n}sk, Poland}

\begin{abstract}
Although `quantum contextuality' is one of the most fundamental non-classical feature, its generic role in information processing and computation is an open quest. In this article, we present a family of distributed computing tasks pertaining to \textit{every} logical proof of Kochen-Specker (KS) contextuality in two different one-way communication scenarios: (I) communication of bounded dimensional system, (II) communication of unbounded dimensional system while keeping certain information oblivious, namely, \textit{oblivious communication} (OC). As the later remains largely unexplored, we introduce a general framework for OC tasks and provide a methodology for obtaining an upper bound on the success of OC tasks in classical communication.  We show that quantum communication comprised of every KS set of vectors outperforms classical communication and perfectly accomplishes the task in both the aforementioned scenarios. We explicitly discuss the communication tasks pertaining to the simplest state independent contextuality sets of dimension three and four. Our results establish an operational significance to single system contextuality and open up the possibility of semi-device independent quantum information processing based on that. Alongside, we identify any advantage in OC tasks as a witness of preparation contextuality. 
\end{abstract}

\maketitle

\section{Introduction} 
The seminal works by Bell-Kochen-Specker \cite{Specker60,Bell66,KS67} demonstrate that the objective reality of sharp values of quantum observables cannot be independent of the measurement context. The Kochen-Specker (KS) reasoning of contextuality stands on the failure of noncontextual assignment of binary values to a set of projectors. However, the general approach to test state independent contextuality (SIC) is based on an inequality consisting of experimentally observed quantities. All noncontextual models satisfy this inequality, while the quantum predictions for any state violate it \cite{Cabello08,Cabello09, YO12, CSW,ana}. Alongside, the notion of \textit{preparation contextuality} \cite{Spekkens2005} denies the context independent objective description of two preparations which are operationally indistinguishable. Despite being one of the most fundamental non-classical features, the generic applicability of quantum contextuality is far from  settled. Many attempts have been made to answer this question in multiple directions. For instance, it is shown that many aspects of quantum computation reveal contextuality \cite{HWVE,anders,Raussendorf}; contextual correlations are also valuable in several information processing  \cite{Spekkens09,KGPLC,OG.etal,Cabello2018,AG.etal,arvind,schmid} and originate novel applications of nonlocal correlations \cite{Brassard05,PC.etal,Abramsky,CLMW,KCK,SR}. While the necessity of quantum contextuality in information processing is one side of the picture, a relevant question to address is whether \textit{every} proof of SIC has any direct inference to quantum advantage in operational tasks. 

This article provides a new perspective to quantum contextuality by showing every  set of vectors that constitutes a logical proof of KS contextuality entails advantage in communication tasks over classical system. We begin by describing a general framework of one-way communication (or distributed computation) task in two distinct scenarios. These two are: (I) \textit{communication in bounded dimension} where the dimension of the communicated system (classical or quantum) is restricted to certain value, and
(II) \textit{oblivious communication} upon the constraint that certain information about the sender's input is unrevealed in communication. As the later remains largely unexplored, we provide a method to obtain optimal bounds on the success probability of oblivious communication (OC) tasks in classical communication.
Moving to the central part of the article we introduce a family of one-way communication tasks, which we refer to as \textit{vertex equality problem}, based on the orthogonal graphs of vector sets with SIC property.  We show that quantum communication comprised of every KS set of vectors outperforms classical communication in both the aforementioned scenarios. While the result holds for any SIC proof involving rank-one projectors in the former scenario, we extend this result for the simplest SIC proof in the later scenario. Significantly, OC does not impose any restriction on the dimension of the communicated system. This implies that even unbounded classical resource cannot reproduce quantum contextual statistics satisfying certain oblivious conditions. We explicitly derive the optimal classical strategies for the vertex equality problem pertaining to Cabello-Estebaranz-GarciaAlcaine (CEG-18) \cite{Cabello99} and Yu-Oh (YO-13) \cite{YO12} vector sets.
Nevertheless, we provide the analytical expression of the optimal success probability in classical communication, applicable to a general vertex equality problem. 
We also study the robustness of quantum communication advantage with respect to white noise and discuss the applicability of quantum contextuality in semi-device independent information processing \cite{qkd,rc}.
Besides, we generalize the observation made in \cite{Spekkens09} that
preparation contextuality is necessary for advantage in OC over classical channel. 

\section{One-way Communication tasks}
A one-way communication task involves a sender (Alice) and a receiver (Bob).
In each round of the task Alice receives an input $x$ from a set $[n_x]:= \{1,\dots,n_x\}$ and sends a message (classical or quantum) to Bob. Bob receives an input $y$ from a set $[n_y]:= \{1,\dots,n_y\}$ and is required to guess some function of their inputs, $f(x,y)$. Bob encodes his answer in an output variable, say, $z$. Let $p(z|x,y)$ represents the probability of obtaining an output $z$ given inputs $x,y$. In this article, we only consider the communication tasks where $f(x,y)$ is binary, i.e., $z\in \{0,1\}$. Furthermore, we deem the figure of merit of the communication problem, i.e., the guessing probability of $f(x,y)$,  as a linear function of the observed probabilities $\{p(z|x,y)\}$. For convenience, one can always normalize the figure of merit such that it takes a value within $[0,1]$. Thus, any linear figure of merit can be expressed as,
\be \label{S}
S =  \sum_{x,y} t(x,y)p(z=f(x,y)|x,y) ,
\ee where $t(x,y)\geq 0, \sum_{x,y}t(x,y)=1$. Here $t(x,y)$ is the normalized weightage for guessing the correct $f(x,y)$. Without loss of generality, we can assume that the inputs $x,y$ are uniformly distributed since the pay-off for non-uniform distribution can effectively be absorbed into $t(x,y)$. This task is trivial if Alice is allowed to send her input $x$ to Bob. The non-triviality and the quantum advantage appear when there is some restriction on the communication from Alice to Bob. Two such distinct scenarios are illustrated below.
\subsection{Communication in bounded dimension}\label{sec:ctbd}
The quantum advantage in communication with bounded dimensional system have been extensively studied within the scope of quantum communication complexity \cite{brassard,wolf,reviewCC} and dimension witness of quantum system \cite{dw1,dw2}. In this scenario, the dimension of the communicated system is bounded by some value, say $d$. In other words, for classical channel the communicated message by Alice (say $\tau$) can be atmost $d$ distinct levels, i.e., $\tau\in \{1,\dots,d\}$. At the receiving end, Bob provides his answer $z$ depending on the received message and his input $y$. Sharing prior classical randomness will not be useful as Alice and Bob seek to maximize a linear function of $\{p(z|x,y)\}$ and any probabilistic strategy is a convex combination of deterministic ones \cite{dw2}. Thus, the general Alice's encoding strategy of $x$ into the message $\tau$ is described by a set of probability distribution $\{p_e(\tau|x)\}$, in which $p_e(\tau|x)$ is the probability of sending a level $\tau \in \{1,\dots,d\}$ for input $x$. On the other hand, we can represent Bob's decoding strategy  by the set of probability distribution $\{p_d(z|y,\tau)\}$ where $p_d(z|y,\tau)$ represents the probability of returning $z$ given his input $y$ and the received message $\tau$. The encoding and decoding strategy should satisfy the normalization conditions, 
\begin{subequations}
\begin{align}
 \forall x, \ &  \sum_{\tau} p_e(\tau|x)=1; \label{nor1} \\
 \forall y,\tau, \ & \ p_d(0|y,\tau)+p_d(1|y,\tau)=1. \label{nor2}
\end{align}
\end{subequations}
Subsequently, the observed probability $p(z|x,y) = \sum_{\tau} p_e(\tau|x) p_d(z|y,\tau)$. 
\begin{obs}\label{obs1}
The optimal value of the figure of merit \eqref{S} in classical communication with bounded dimension $d$, denoted by $S^{(1)}_c$, can be obtained only in terms of encoding strategy $\{p_e(\tau|x)\}$ as follows,
\beq \label{ints}
&S^{(1)}_c = & \max\limits_{\{p_e(\tau|x)\}} \sum\limits_{y} \sum\limits_{\tau} \max \bigg( \sum\limits_{x\in F^0_y} t(x,y) p_e(\tau|x), \nonumber \\
& & \qquad \qquad  \sum\limits_{x\in F^1_y} t(x,y) p_e(\tau|x) \bigg),
\eeq
where $\tau\in \{1,\dots,d\}$ and $F^z_y$ is a subset of $[n_x]$ such that $f(x,y)=z$.
\end{obs}
\begin{proof}
For each $y$, let us define a set $F^z_y \subset [n_x]$ such that $f(x,y)=z$.
We can express $S^{(1)}_c$ by splitting the summation over $x$ into two sets $F^0_y,F^1_y$ as follows,
\begin{widetext}
\beq \label{calints}
& S^{(1)}_c \ & = \max_{\{p_e(\tau|x)\},\{p_d(z|y,\tau)\}}
\sum_y \bigg( \sum_{x\in F^0_y} t(x,y) p(z=0|x,y) + \sum_{x\in F^1_y} t(x,y) p(z=1|x,y) \bigg) \nonumber \\
&& =   \max_{\{p_e(\tau|x)\},\{p_d(z|y,\tau)\}} \sum_{y} \sum_{\tau} \left( \left(\sum_{x\in F^0_y} t(x,y) p_e(\tau|x)\right) p_d(0|y,\tau)
 +  \left(\sum_{ x\in F^1_y} t(x,y) p_e(\tau|x)\right) p_d(1|y,\tau) \right). 
\eeq
\end{widetext} 
Due to the condition \eqref{nor2}, the above expression simplifies to \eqref{ints} and the optimal decoding is given by, \\
 \be \label{dsmain}
p_d(0|y,\tau) =
  \begin{cases}
    1, \text{if} \sum\limits_{x\in F^0_y} t(x,y)  p_e(\tau|x) \geq \sum\limits_{x\in F^1_y} t(x,y)  p_e(\tau|x)\\
    0,\text{if} \sum\limits_{x\in F^0_y} t(x,y)p_e(\tau|x) < \sum\limits_{x\in F^1_y} t(x,y)  p_e(\tau|x).
  \end{cases}
\ee
The above observation implies, given any encoding strategy \eqref{dsmain} the optimal decoding strategy for Bob is fixed and deterministic.
\end{proof}
We seek to optimize of the right-hand-side of \eqref{ints} under the constraints \eqref{nor1}. Obviously, the variables $\{p_e(\tau|x)\}$ form a polytope whose extremal points are deterministic, i.e., $p(\tau|x)\in \{0,1\}$. 
Since the `$\max$ function' of two linear functions is a convex function, the optimal value of the right-hand-side of \eqref{ints} can be obtained by evaluating the expression at all possible deterministic encoding strategies.

Whereas, in the case of quantum channel, the communicated quantum state for input $x$, say $\rho_x$, should belong to $d$-dimensional Hilbert space. And subject to input $y$, Bob performs a measurement $\{M^z_y\}$ on the communicated system and returns the measurement outcome $z$. This quantum strategy yields $S = \sum_{x,y} t(x,y) \tr(\rho_x M^{z=f(x,y)}_y)$.

\subsection{Oblivious Communication}\label{sec:oc}
\begin{figure}[H]
\centering
\includegraphics[scale=0.63]{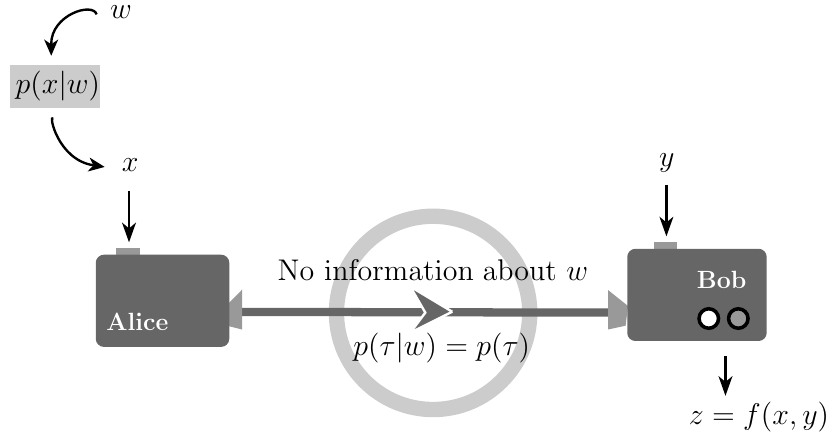}
\caption{\textit{Oblivious communication problem} between a sender (Alice) and a receiver (Bob) who receive inputs $x$ and $y$ respectively. Alice's input $x$ comes through a local channel whose input variable is $w$. In classical communication, Alice communicates arbitrary large number of classical messages $\tau$ to Bob which do not contain any information about the oblivious variable $w$. The goal is to guess $f(x,y)$ with maximum probability.}
\label{fig7}
\end{figure}

On the contrary with the previous scenario, OC task with quantum resources is mostly unexplored. In OC, there is no restriction on the dimension of the communicated system. Instead, we impose secrecy of certain information in the communication. Let us first propose a general framework of one-way OC task (see Fig. \ref{fig7}). Here, Alice's input $x$ comes through a local channel described by the conditional probability $p(x|w)$ where $w \in [n_w]:= \{1,\dots,n_w\}$ is the input variable of that channel. The communication from Alice to Bob is unbounded. The only condition is that the information about the \textit{oblivious variable} $w$ should not be revealed in the communication including Bob.

Formally, the oblivious condition in classical communication implies Alice's encoding strategy $\{p_e(\tau|x)\}$ should satisfy the condition that $p_e(\tau|w)$ is independent of $w$:
\beq \label{goc}
&& \forall \tau, \ \forall w,w' \in [n_w], \nonumber  \\
&& p_e(\tau|w)=\sum_x p(x|w) p_e(\tau|x) = p_e(\tau|w') .\eeq
Additionally, the dimension of the classical message is not bounded, i.e., $\tau \in \{1,\dots,N\}$ where $N$ can be arbitrarily large. If Alice and Bob share a classical random variable $r$, the oblivious constraint demands, $\forall \tau,r,\ p_e(\tau|w,r)=p_e(\tau|r)$, that is, the variable $w$ should be oblivious to the receiver Bob given the shared randomness too. Using the fact that the input variable $w$ is independent of $r$, one obtains, $\forall \tau,r, \ p_e(\tau,r|w) = p_e(\tau,r)$. As the dimension of the communicated message can be arbitrarily large, without loss of generality, one can include the shared randomness into a larger message $\tau'=(\tau,r)$.  Thus, it suffices to consider the classical strategies without shared randomness. This fact also suggests that whenever Alice's encoding strategy is probabilistic, i.e., $p_e(\tau|x) \notin \{0,1\}$, it  should be realized using Alice's local randomness.
Note that, Alice has information about $w$ from her input $x$, but the primary goal is to encode the input $x$ efficiently in such a way that information about $w$ should be utterly oblivious to any other party who does not have access to Alice's lab.  

It is non-trivial to obtain the classical optimal success probability \eqref{S} in OC tasks since $\tau$ can take arbitrary large number of values.  Moreover, the general encoding strategy $\{p_e(\tau|x)\}$ may not be deterministic. Although any probabilistic encoding strategy which satisfies the oblivious constraints \eqref{goc} is a convex combination of deterministic strategies, but those deterministic strategies might not satisfy the oblivious constraints \eqref{goc}. We will see later that the optimal encoding schemes are indeed not necessarily deterministic.  Below we provide a method to obtain an upper bound on the success probability.

\begin{prop} \label{prop:sc}
An upper bound on the optimal value of the figure of merit \eqref{S} in classical communication under oblivious constraints \eqref{goc}, denoted by $S^{(2)}_c$, can be obtained as follows,
\be \label{Msimexpmain}
S^{(2)}_c \leq \max_{\{q_x\}} \sum_{y} \max \left(\sum\limits_{x\in F^0_y} t(x,y) q_x,\sum\limits_{x\in F^1_y} t(x,y) q_x  \right),
\ee
where the maximization is taken over $n_x$ number of variables $q_x$ that satisfy the conditions,
\beq \label{Mc3}
&\forall x\in [n_x],\ q_x \geq 0;\ \forall w,\ \sum_{x} p(x|w) q_x = 1.
\eeq
\end{prop}
\begin{proof}
Following the previous argument in \textit{Observation} \ref{obs1}, the optimal classical value of $S^{(2)}_c$ can be simplified only in terms of encoding strategy,
\beq \label{intsoc}
 S^{(2)}_c &=&  \max_{\{p_e(\tau|x)\}} \sum_{y} \sum_{\tau} \max \bigg( \sum_{x\in F^0_y} t(x,y) p_e(\tau|x), \nonumber \\
 && \sum_{x\in F^1_y} t(x,y) p_e(\tau|x) \bigg).
\eeq
In contrast with the previous scenario, we seek to maximize  the right-hand-side of \eqref{intsoc}  under the conditions: $(i)$ $\tau$ can take arbitrary large number of values, $(ii)$ the oblivious constraints \eqref{goc} and normalization,
\begin{subequations}
\begin{align}
& \forall x, \ \sum_{\tau} p_e(\tau|x) =1, \label{c1} \\
& \forall \tau, w, \ \ p_e(\tau):= p_e(\tau|w) = \sum\limits_{x} p(x|w) p_e(\tau|x) .  \label{c2}
\end{align}
\end{subequations}
Since $\tau$ can take arbitrary number of values, the maximization in \eqref{intsoc} is taken over arbitrarily large number of variables which is not feasible. 
By defining another variable 
\be 
q^{\tau}_x = \frac{p_e(\tau|x)}{p_e(\tau)},
\ee 
where $p(\tau)$ is defined in \eqref{c2}, and substituting $ p_e(\tau|x)=p_e(\tau)q_x^{\tau}$ in \eqref{intsoc} we express $S_c^{(2)}$ as follows,
\be \label{s2}
S^{(2)}_c = \max_{\{q^{\tau}_x\}} \sum_{\tau} p_e(\tau) \underbrace{ \bigg( \sum_{y} \max [\sum\limits_{x\in F^0_y} t(x,y) q^\tau_x,\sum\limits_{x\in F^1_y} t(x,y) q^\tau_x ]  \bigg) }_{\chi_e(\tau)}.
\ee
Besides, with the aid of \eqref{c1} and \eqref{c2} we have,
\be \label{pe}
\sum_{\tau} p_e(\tau) = \sum_x p(x|w) \left(\sum_{\tau} p_e(\tau|x) \right) = 1. \ee
Hence, \eqref{s2} can be interpreted as a convex combination of $\chi_e(\tau)$ with coefficient $p_e(\tau)$. This observation leads to the fact that $S^{(2)}_c$ is bounded by the maximum possible value of $\chi_e(\tau)$, i.e.,
\be \label{simexpmain}
S^{(2)}_c \leq \max_{\{q_x\}} \sum_{y} \max \left(\sum\limits_{x\in F^0_y} t(x,y) q_x,\sum\limits_{x\in F^1_y} t(x,y) q_x  \right).
\ee
Note that, the above expression is independent of the number of $\tau$. Another way to interpret \eqref{simexpmain} is, $S^{(2)}_c$ is upper bounded by the success probability pertaining to a single message $\tau$. So, it is convenient to denote the variable $q^{\tau}_x$ by $q_x$ in \eqref{simexpmain}. By dividing $p_e(\tau)$ on both sides of \eqref{c2}, we see that $q_x$ satisfies the following constraints,
\beq \label{c3}
&\forall x \in [n_x],\ q_x \geq 0;\quad \forall w,\ \sum_{x} p(x|w) q_x = 1.
\eeq
Consequently, the upper bound on $S^{(2)}_c$ reduces to the maximum value of the right-hand-side of \eqref{simexpmain} under the constraints \eqref{c3}. In a nutshell, this method simplifies the optimization problem of arbitrarily large number of variables to $n_x$ number of variables. It is worth noting that \eqref{simexpmain} provides an upper bound on $S^{(2)}_c$ which might not be tight since \eqref{c2} has not been imposed for all $\tau$. \\
The variables $q_x$ satisfying \eqref{c3} form a polytope. The optimal value of the right-hand-side of \eqref{simexpmain} can be obtained by evaluating the expression at these extremal points of the polytope. Again, it follows due to the fact that the `$\max$ function' of two linear functions is a convex function. 
\end{proof}

In the case of quantum channel, for input $x$ Alice sends a quantum state $\rho_x$ which belongs to a Hilbert space of any dimension.  The oblivious condition demands the effective quantum state for an oblivious variable $w$ (say $\rho_w$) is same for all $w$, i.e., 
\be \label{quantumoc}
\forall w,w'\in [n_w], \  \rho_w = \sum_x p(x|w)\rho_x = \rho_{w'}. \ee
We remark that the OC problem is a generalization of oblivious transfer which serves as a primitive for several classical and quantum cryptographic protocols \cite{crypto,qcrypto,qcrypto1,marcin} and privacy-preserving computation \cite{ppc}. 
It has been shown that \textit{preparation contextuality} is useful in parity oblivious random access code \cite{Spekkens09,banik,AC.etal,AA.etal,AH.etal,pan}. Parity oblivious multiplexing can be interpreted as a particular case of the above general framework, wherein the oblivious variable $w$, that is, the parity of a set of input bits, is a function of the input $x$. While the approach to obtain the classical bound in \cite{Spekkens09} is applicable to random access code, the proposed method is universal. 
We also generalize the ontological implication \cite{Spekkens09} that \textit{any advantage in OC reveals preparation contextuality}. In other words, the classical bound for an OC task is also the optimal bound in all theories that satisfy the notion of preparation noncontextuality \cite{Spekkens2005}. The proof of this fact is discussed in Section \ref{sec:pc}. It is also noteworthy that there exists a generic relation of quantum advantage between these two communication scenarios \cite{saha}.

\section{Vertex equality problem}
We now present a class of communication tasks pertaining to SIC proofs. Remark that the logical proofs of KS contextuality are particular instances of SIC \cite{Cabello08,Cabello09}. We only consider SIC proofs that comprise a set of rank-one projectors which we refer to \textit{SIC set} (or \textit{KS set} in particular).   

\begin{definition}[SIC graph]\label{def:graph} Pertaining to every SIC set of rank-one projectors, we define a graph $G$ where each vertex corresponds to a projector (or equivalently a vector), and two vertices are adjacent if the corresponding projectors (or vectors) are orthogonal. This is known as orthogonal graph \cite{CSW}.
\end{definition}

We denote the vertices by $v\in \{1,\dots,|G|\}$ and the total number of vertices, i.e., the order of graph $G$, by $|G|$.  The \textit{neighborhood} of a vertex $v$, denoted by $N_v$, is the induced subgraph of $G$ consisting all the adjacent vertices of $v$. We refer the vector that corresponds to the vertex $v$ of the SIC graph by $|v\rangle$. Thus, $|N_v|$ is the number of orthogonal vectors to $|v\rangle$ present in the SIC set.

\begin{definition}[Vertex equality problem] \label{def:vep}
Given a SIC graph $G$, Alice and Bob receive input from the vertex set of $G$, i.e., $x,y \in \{1,\dots,|G|\}$, thereby $n_x=n_y=|G|$. Whenever $y\in \{x,N_x\}$, Bob's aim is to guess the function,
\be 
 f(x,y) = \begin{cases}
0, \ \text{ if } \ y=x\\
1, \ \text{ if } \ y\in N_x.
\end{cases}
\ee 
In other words, they seek to maximize the guessing probability of  whether $x=y$ or $x\neq y$ conveyed by the output $z=0$ or 1 respectively.
\end{definition} 

Thus, in the vertex equality problem essentially we are only interested in those runs when Bob's input $y$ is connected or equal to Alice's input $x$ in the graph. In other words, $t(x,y)$ (in the figure of merit \eqref{S}) is \textit{non-zero} if and only if $y \in \{N_x,x\}$. 
As an equality problem where the inputs belong to the vertices of SIC graph, we call this task \textit{vertex equality problem}. Below, we define a quantum communication protocol of this task.

\begin{definition}[Quantum SIC strategy]\label{def:strategy} Upon receiving an input $x$ Alice prepares the quantum state $\rho_x = |x\rangle \langle x|$ where $|x\rangle$ corresponds to the vector associated with the vertex $x$ in the SIC graph, and sends to Bob. For input $y$ Bob performs a binary outcome measurement $\{M^z_y\} := \{M^0_y,M^1_y\}$ on the received system $\rho_x$, where $M^0_y=|y\rangle \langle y|, M^1_y=\mathbbm{1} - |y\rangle \langle y|$.
\end{definition}

\subsection{Vertex equality problem in bounded dimension}
\begin{figure}[H]
\centering
\includegraphics[scale=0.7]{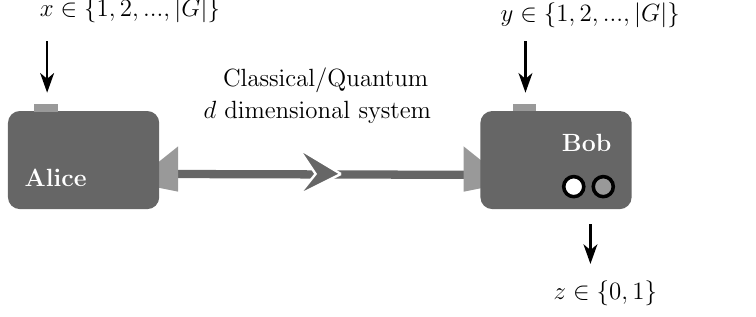}
\caption{Vertex equality problem in bounded dimension. Alice receives an input $x$ from the vertex set of SIC graph $G$ and communicates a system to Bob. Bob obtains input $y$ from the same set and returns a binary outcome to guess if $x=y$ or not. The relevant runs are those where $y\in \{N_x,x\}$. The dimension of communicated system is bounded by $d$, where $d$ is the minimum dimension of the Hilbert space in which the SIC set is realized.}
\label{fig1}
\end{figure}

First, we consider the scenario (see Fig.\ref{fig1}) restricting the dimension of the transmitted system (classical or quantum) by Alice. The upper bound on the dimension, that is referred to $d$, is equal to the dimension of the Hilbert space in which the SIC set of vectors is realized. For simplicity, we take $t(x,y)$ to be uniform, and hence we seek to maximize,
\beq \label{sbd}
&S^{(1)} = \sum\limits_{\substack{x,y \\ x=y}} t(x,y) p(0|x,y)
+  \sum\limits_{\substack{x,y \\ y\in N_x}} t(x,y) p(1|x,y), \nonumber \\
&\text{where } \ t(x,y) = \frac{1}{N}, \ N = \sum_x |N_x| + |G|.
\eeq
\begin{prop}\label{prop:sbd}
For the vertex equality problem in bounded dimension, the maximum value of $S^{(1)}$ \eqref{sbd}, i.e. one, can be achieved in quantum communication. While for the classical communication,
\be \label{cb}
S^{(1)}_c = 1 - \frac{\kappa}{N},
\ee which is strictly less than one. Here $\kappa$ is the minimum number of vertices which are not properly colored (i.e., at least one of its neighbor is assigned the same color) when maximum $d$ colors are used to color all the vertices in $G$. 
\end{prop}
\begin{proof}
It is clear from the construction of the Quantum SIC strategy \eqref{def:strategy} that when $x=y$, $p(0|x,y)=1$, and whenever $y\in N_x$, since $\rho_x$ has full support in $\mathbbm{1} - |y\rangle \langle y|$, $p(1|x,y)=1$. Thus, quantum SIC strategy \eqref{def:strategy} provides the correct answer for all relevant inputs. Moreover, this strategy involves sending only $d$-dimensional quantum system which fulfils the requirement of bounded dimension in communication. \\
Subject to classical communication, following \eqref{ints} we know,
\be \label{ght}
S^{(1)}_c =  \max_{\{p_e(\tau|x)\}} \frac{1}{N} \sum^{|G|}_{y=1} \sum^d_{\tau=1}  \max \bigg( p_e(\tau|x=y), \sum_{ x\in N_y} p_e(\tau|x) \bigg).
\ee  Further using the facts that $\max(a,b)=a+b-\min(a,b)$ and $\sum_{x,y} \sum_{\tau} p_e(\tau|x) = N$, one may re-express \eqref{ght} as,
\beq \label{c4}
&S^{(1)}_c = 1 -  \min\limits_{\{p_e(\tau|x)\}} \frac{1}{N} \sum^{|G|}\limits_{y=1} \sum\limits^d_{\tau=1}  \min & \bigg( p_e(\tau|x=y), \nonumber \\
&& \sum_{ x\in N_y} p_e(\tau|x) \bigg).
\eeq
Without loss of generality, we can assume that there exists a deterministic encoding strategy $\{p_e(\tau|x)\}$ that yields $S^{(1)}_c$. That is, Alice sends one of $d$ different levels for each input $x$. Therefore, it is equivalent to assign one of the $d$ colors to each vertex of $G$. From \eqref{c4}, we see that $S^{(1)}_c=1$ holds when
\be \label{sss1}
\forall y, \ \sum_{\tau} \min \left( p_e(\tau|x=y), \sum_{ x\in N_y} p_e(\tau|x) \right)=0 .
\ee  The above condition is satisfied if and only if the colors assigned to all the adjacent vertices of $y$ are different from the color assigned to $y$. We know that, since the chromatic number of any SIC graph pertaining to any SIC realization of rank-one projectors is strictly greater than $d$ \cite{Cabello11,RH} (Theorem 4 in  \cite{Cabello11}), it is not possible to perfectly color all the vertices of a SIC graph with $d$ colors such that no two adjacent vertices are assigned the same color. Hence, the value of $S^{(1)}_c$ is strictly less than one. In fact, this is true for non-uniform values of $t(x,y)$. Now, given a vertex coloring, the left-hand-side of \eqref{sss1} is 1 if the vertex $y$ is not properly colored, i.e., at least one of its neighbor is assigned the same color.
Thus, $\min\limits_{\{p_e(\tau|x)\}} \sum_{y} \sum_{\tau}  \min \bigg( p_e(\tau|x=y), \sum_{ x\in N_y} p_e(\tau|x) \bigg)$ is equal to the minimum number of vertices which are not properly colored when maximum $d$ colors are used to color all the vertices in $G$, obtaining the bound given by \eqref{cb}.
\end{proof}
\begin{figure}[H]
\centering
\includegraphics[scale=0.7]{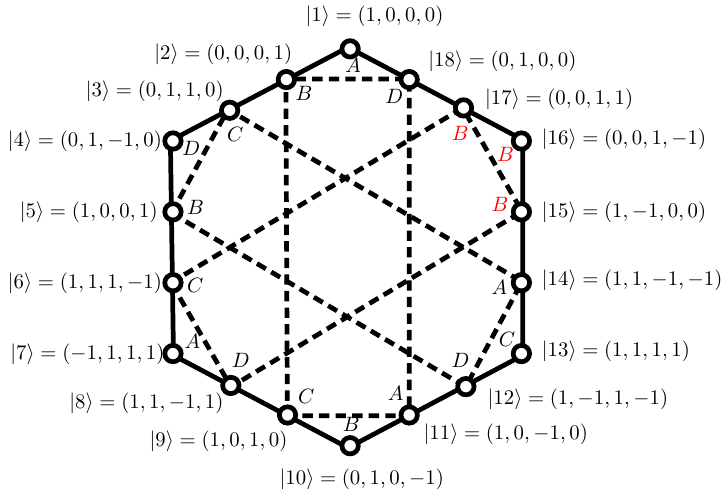}
\caption{\textit{CEG-18 set.} A simplified version of the orthogonal graph of CEG-18 set of vectors. Each vertex corresponds to a vector. Six edges of the heptagon and the three dashed rectangles connecting four vertices form the basis ($d$-clique). Each vector belongs to two different basis and orthogonal to six other vectors. The optimal classical strategy is also shown by assigning four distinct levels $\{A,B,C,D\}$ to all vertices. The assigned colors to the vertices $15,16$ and $17$ are already used for one of their adjacent vertices. Thus, there are three vertices which are not properly colored.}
\label{fig2}
\end{figure}
\begin{coro}
Any SIC set of rank-one projectors in dimension $d$ provides device-independent witness \cite{dw1,dw2} of $d$-dimensional quantum system. 
\end{coro}

One may quantify the quantum advantage as the difference between the dimensions of the system required to communicate in classical and quantum communication to accomplish vertex equality task perfectly. For classical system, the minimum dimension required to achieve $S^{(1)}=1$ is the chromatic number of the graph which is greater than $d$ for any SIC graph. It is also noteworthy that, \textit{Proposition} \ref{prop:sbd} is valid for any graph whose  chromatic number is higher than the minimum dimension of Hilbert space required to realize the graph. 

\begin{figure}[H]
\centering
\includegraphics[scale=0.4]{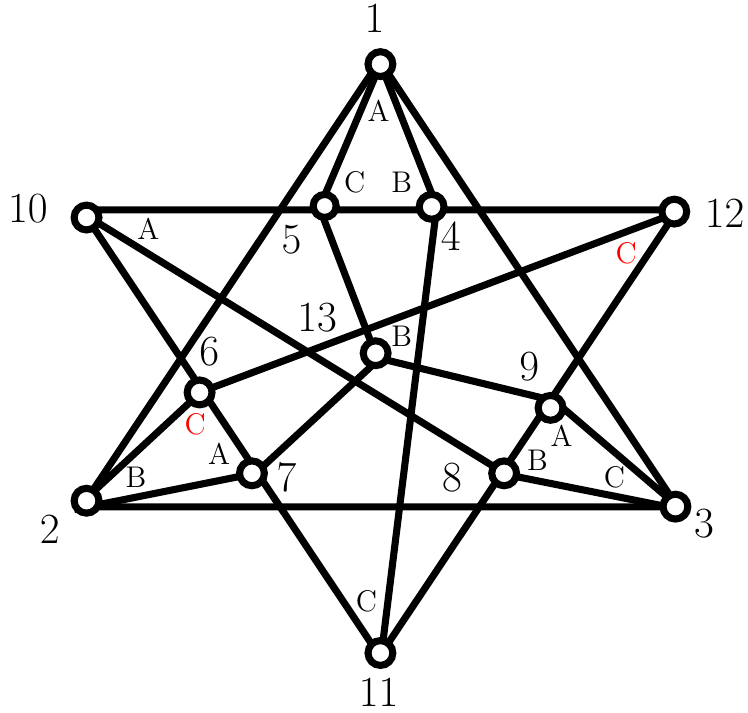}
\caption{\textit{YO-13 set.} The orthogonal graph of YO-13 set of vectors. The explicit expression of the unnormalized vectors are as follows, $\ket{1}=(1,0,0),\ket{2}=(0,1,0),$$\ket{3}=(0,0,1),$$\ket{4}=(0,1,1),$$\ket{5}=(0,1,-1),$$\ket{6}=(1,0,1),$$\ket{7}=(1,0,-1),$$\ket{8}=(1,1,0),$$\ket{9}=(1,-1,0),$$\ket{10}=(-1,1,1),$$\ket{11}=(1,-1,1),$$\ket{12}=(1,1,-1),$$\ket{13}=(1,1,1)$. The optimal classical strategy is also shown by assigning three distinct levels $\{A,B,C\}$ to all vertices. There are two improperly colored vertices $6,12$.}
\label{fig6}
\end{figure}
The SIC graphs of CEG-18 \cite{Cabello99} and YO-13 \cite{YO12} sets of vectors are presented in Figure \ref{fig2} and \ref{fig6}. It can be checked  that the independence number of CEG-18 graph is four. Thus with four colors, maximum 16 out of 18 vertices can be properly colored and thus, $\kappa $ cannot be less than 3.  In Fig. \ref{fig2} a graph coloring is shown with $\kappa =3$. While for YO-13 graph, $\kappa=2$  which is the minimum for any SIC graph (Fig. \ref{fig6}).

\subsection{Vertex equality problem in oblivious communication}
Let's consider a particular instance of the general OC task described in Sec \ref{sec:oc}. For each $w\in \{1,\dots,n_w\}$, there exists a set $\Omega_w \subset  [n_x]$ such that 
\be 
p(x|w) = 
\begin{cases} 
\frac{1}{|\Omega_w|}, \text{ if } \ x \in \Omega_w \\
0, \text{ otherwise,} 
\end{cases}
\ee 
where $|\Omega_w|$ as the cardinality of $\Omega_w$. Consequently, the implicit constraints in classical \eqref{goc} and quantum \eqref{quantumoc} communication are as follows:
\beq \label{Moc}
&&\forall \tau, \ p_e(\tau) = p_e(\tau|w) = \frac{1}{|\Omega_1|}\sum_{x\in \Omega_1} p_e(\tau|x) = \dots \nonumber \\
&& \qquad \qquad \qquad \quad \dots = \frac{1}{|\Omega_{n_w}|} \sum_{x\in \Omega_{n_w}} p_e(\tau|x); \nonumber \\
&&\rho_w = \frac{1}{|\Omega_1|} \sum_{x\in \Omega_1} \rho_x = \dots = \frac{1}{|\Omega_{n_w}|} \sum_{x\in \Omega_{n_w}} \rho_x .
\eeq
These constraints imply the effective communicated state when $x$ belongs to the set $\Omega_w$ is indifferent for all $w$. Therefore, Alice does not reveal information about those subsets $\Omega_w$ to which her input $x$ belongs. Here we consider the same vertex equality problem under such oblivious constraints taking $\Omega_w$ are the maximum cliques of the SIC graph.

The clique of a graph is a subset of vertices such that each pair of them are connected by edge. If a SIC set of vectors is realized in dimension $d$, then the maximum clique of the SIC graph, i.e., a clique with the maximum number of vertices, has no more than $d$ vertices. If some vertices do not belong to maximum clique (or $d$-clique), then we consider extended SIC graph $G^e$ with additional vertices such that each vertex belongs to at least one $d$-clique. In other words, we include additional vectors in SIC set such that each vector belongs to some basis. Same as the vertex equality task, here Alice and Bob receive input $x$ and $y$, respectively, from the vertex set $\{1,2,...,|G^e|\}$ and want to guess whether $x$ and $y$ are the same or not whenever $y\in \{x,N_y\}$. The oblivious variable $w$ corresponds to a set of vertices $\Omega_w \subset \{1,2,...,|G^e|\}$ that forms $d$-cliques. Thus, each $\Omega_w$ representing a $d$-clique of $G^e$ contains $d$ elements, and $n_w$ is the total number of $d$-cliques in $G^e$. Such oblivious constraints, referred to as \textit{clique-oblivious condition}, implies no information is communicated about the maximum clique ($\Omega_w$) in which the input $x$ belongs.

Let us consider the example of CEG-18 set.
In the CEG-18 graph (Fig. \ref{fig2}), each vertex belongs to exactly two maximum cliques, so we do not need to consider an extended graph. The nine maximum cliques (see Fig. \ref{fig2}) correspond to the oblivious variable $w \in \{1,\dots,9\}$ are,
\beq \label{18c}
&\Omega_1 =\{1,2,3,4\}, \Omega_2 =\{4,5,6,7\}, \Omega_3 =\{7,8,9,10\}, \nonumber \\
& \Omega_4 =\{10,11,12,13\}, \Omega_5 =\{13,14,15,16\}, \nonumber \\
& \Omega_6 =\{16,17,18,1\},  \Omega_7 =\{2,9,11,18\},\nonumber \\
& \Omega_8 =\{3,5,12,14\}, \Omega_{9} =\{6,8,15,17\}.
\eeq 
The clique-oblivious condition \eqref{Moc} only allows encoding strategies such that for all $\tau,$ the quantity $\frac{1}{4}\sum_{x\in \Omega_w} p_e(\tau|x)$ is same for every $w$. Note that the trivial strategy where Alice communicates her input $x$, that is $p_e(\tau|x)=\delta_{\tau,x}$, does not satisfy this oblivious condition since $p_e(\tau=1|w=1)=1/4$ while $p_e(\tau=1|w=2)=0$.

In general, under clique-oblivious conditions Alice and Bob seek to maximize the following quantity,
\beq \label{soc}
& S^{(2)} = \sum\limits_{\substack{x,y \\ x=y}} t(x,y) p(0|x,y)
+  \sum\limits_{\substack{x,y \\ y\in N_x}} t(x,y) p(1|x,y), \nonumber \\
& \text{taking } t(x,y) = \frac{c_x}{N}, \ N = \sum\limits^{n_w}_{w=1} |\Omega_w|.
\eeq
Here $c_x=m_x/(|N_x|+1)$ where $m_x$ is the number of $d$-cliques ($\Omega_w$'s) in which $x$ appears and $|N_x|$ is number of adjacent vertices of $x$. For any KS set the cardinality of maximum clique $|\Omega_w|$ is equal to $d$. Later, for YO-13 set, we will consider additional $\Omega$ of different cardinality than $d$. Yet again, $y \in \{N_x,x\}$ are the relevant runs that appear in the figure of merit. However, $t(x,y)$ is chosen to be non-uniform here. 
The upper bound derived for a general OC tasks in \eqref{simexpmain}-\eqref{c3} simplifies for the vertex equality problem \eqref{soc} to,
\beq \label{simexp}
 S^{(2)}_c \leq   \frac{1}{N} \sum_{y}  \max \bigg( c_{y} q_y, \sum_{ x\in N_y} c_x q_x, \bigg)
\eeq 
where the variables $q_x$ satisfy the following conditions,
\be \label{cq}
\forall x, q_x \geq 0; \ \ \forall w, \ \frac{1}{|\Omega_w|} \sum_{x\in \Omega_w} q_x = 1.
\ee
\begin{figure}[H]
\centering
\includegraphics[scale=0.63]{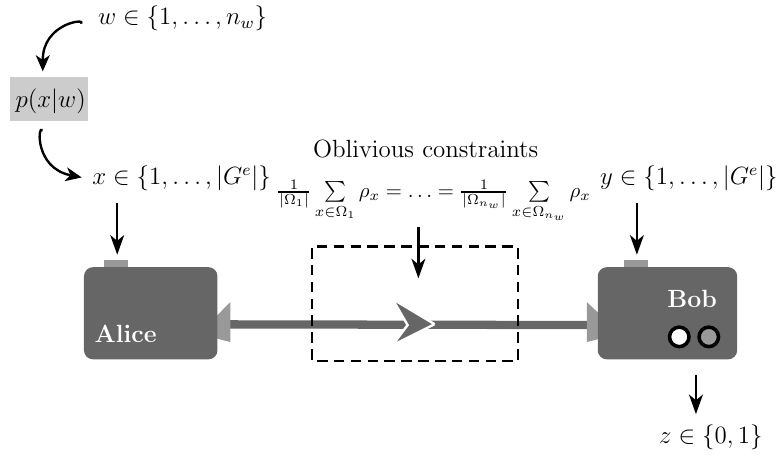}
\caption{Vertex equality problem in oblivious communication based on extended SIC graph. Here $\Omega_w$ denote the maximum cliques of the graph. The communicated system is required to satisfy the clique-oblivious constraints \eqref{Moc}. }
\label{fig3}
\end{figure}
\begin{prop}\label{prop:soc}
For the vertex equality problem based on any extended KS graph under clique-oblivious conditions \eqref{Moc}, quantum communication outperforms its classical counterpart. Particularly, the value of $S^{(2)}$ \eqref{soc} is strictly less than one in classical communication, whereas it is one in quantum communication.
\end{prop}
\begin{proof}
It can be readily verified that Quantum SIC strategy \eqref{def:strategy} satisfies the clique-oblivious conditions \eqref{Moc} and produces the correct answer for all relevant inputs. By the construction of extended SIC graph, each maximum clique corresponds to a basis. This implies, for all $w$, $\rho_w$ is the maximally mixed state in dimension $d$. \\
For classical channel, let Alice sends a level $\tau$ for input $x$ with nonzero probability,  i.e., $q_x = p_e(\tau|x)/p(\tau)>0$. If there exists $x'\in N_x$, such that $q_{x'}>0$, then either for the input $(x,x')$ or the input $(x',x)$, Bob guesses the wrong answer with non-zero probability. Let us provide a rigorous explanation of this fact. Due to \eqref{soc} and \eqref{cq}, we obtain
\be 
\sum_x m_x q_x = \sum_w \sum_{x\in \Omega_w} q_x = \sum_w |\Omega_w| =N.\ee
Using the above relation and replacing the `max' function by `min' [$\max(a,b) = a+b-\min(a,b)$], \eqref{simexp} can be re-expressed as follows,
\beq
&&S^{(2)}_c \leq   \frac{1}{N} \sum_{y}  \max \bigg( c_y q_y, \sum_{ x\in N_y} c_x q_x, \bigg) \nonumber \\
&& = \frac{1}{N} \sum_{y}  \bigg( c_{y} q_y + \sum_{ x\in N_y} c_x q_x -  \min \bigg( c_y q_y, \sum_{ x\in N_y} c_x q_x \bigg)\bigg) \nonumber \\
&&= \frac{1}{N} \sum_{y} m_y q_y - \frac{1}{N} \sum_{y}  \min \bigg( c_y q_y, \sum_{ x\in N_y} c_x q_x \bigg) \nonumber \\
&& = 1 - \frac{1}{N} \sum_{y}  \min \bigg( c_y q_y, \sum_{ x\in N_y} c_x q_x \bigg).
\eeq  Thus, the task is perfectly accomplished only if
\be \label{sss2}
\forall y, \ \min \left( c_y q_y, \sum_{ x\in N_y} c_x q_x \right)=0 .
\ee 
Let's recall that $q_x = p_e(\tau|x)/p(\tau)$, and consequently \eqref{sss2} implies either $p_e(\tau|y)=0$, or for all $x\in N_y,\ p_e(\tau|x)=0$.
This leads to the following observations. \\
\textit{Fact 1.} If $p_e(\tau|x)>0$, then \eqref{sss2} implies for all other inputs belong to $N_{x}$, Alice cannot communicate $\tau$. \\
\textit{Fact 2.} The clique-oblivious condition \eqref{Moc} necessitates for each $d$-clique $\Omega_w$ there exists a vertex $x\in \Omega_w$ such that $p_e(\tau|x)>0$. \\
Consider an assignment of `1' or `0' to each vertex $x$ if $p_e(\tau|x)>0$ or $p_e(\tau|x)=0$, respectively. Then \textit{Fact 1-2} suggests that we can assign `0' or `1' values to all vectors such that no two orthogonal vectors are assigned value `1' and exactly one vector in each basis is assigned `1'. From the very definition of KS set, this implies a contradiction. Therefore, $\sum_{y}  \min \bigg( c_y q_y, \sum_{ x\in N_y} c_x q_x \bigg)$ is non-zero, and $S^{(2)}_c$ is strictly less than one.
\end{proof}

\subsection*{CEG-18 set}
\begin{figure}[H]
\centering
\includegraphics[scale=0.7]{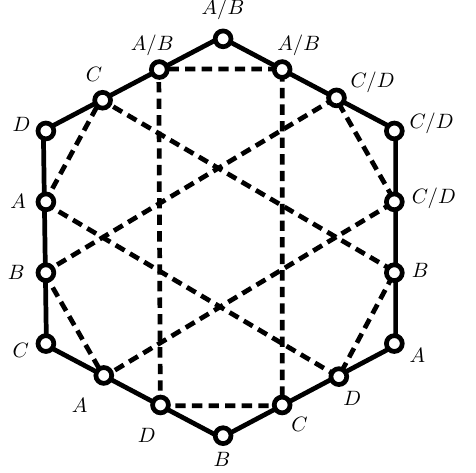}
\caption{The optimal classical encoding strategy in the clique-oblivious vertex equality task pertaining to CEG-18 is shown. The strategy comprises of four distinct levels $\tau \in \{A,B,C,D\}$. $A/B$ means Alice sends an equal mixture of levels $A$ and $B$. One can check that, $\forall \tau, w, \ p_e(\tau|w)=1/4$, and thus each level satisfies the oblivious constraints \eqref{Moc}. 
Taking into account the optimal decoding \eqref{dsmain} one can verify $S^{(2)}_c=20/21$ for this particular strategy.}
\label{fig4}
\end{figure}

\begin{prop} \label{prop:18}
The optimal classical value $S^{(2)}_c$ is 20/21 for the clique-oblivious vertex equality problem based on CEG-18 set.
\end{prop} 
\begin{proof}
Due to the elegant symmetry of CEG-18 set, we have $c_x=2/7$ for all $x\in \{1,\dots,18\}$ and $N=36$. Subsequently, \eqref{simexp} can be stated as,
\beq 
S^{(2)}_c &\leq& \frac{1}{126} \sum_{y} \max \left(q_y, \sum_{ x\in N_y} q_x \right) \nonumber \\
&=& \frac{1}{126} \sum_{y} \max \left(q_y, 8 - 2q_y \right)  ,
\eeq
where we use the fact that 
\be 
\forall x, \ \sum_{ x\in N_y} q_x = 2\sum_{x\in \Omega} q_x - 2q_y = 8 - 2q_y. \ee
It follows from \eqref{cq} that due to the oblivious conditions \eqref{18c}, $q_x$ satisfies the following constraints,
\beq
&\forall x, q_x \geq 0; ~ q_1+q_2+q_3+q_4=q_4+q_5+q_6+q_7 \nonumber \\ \nonumber
& =q_7+q_8+q_9+q_{10}  =q_{10}+q_{11}+q_{12}+q_{13} \\ \nonumber
& = q_{13}+q_{14}+q_{15}+q_{16} = q_{16}+q_{17}+q_{18}+q_{1} \\ 
&=q_{2}+q_{18}+q_{9}+q_{11}=q_{3}+q_{5}+q_{12}+q_{14} \nonumber \\ 
& =q_{6}+q_{8}+q_{15}+q_{17} = 4.
\eeq
One obtains the extremal points of the polytope of $q_x$ by simple linear programing. There are 146 extremal points. 
Following the argument given before, that the `$\max$' function of two linear functions is a convex function, the optimal value of the quantity $ \sum_{x} \max [q_x, 8 - 2q_x]$ can be obtained from these extremal points of the polytope. So, one may easily retrieve the maximum value of $ \sum_{x} \max [q_x, 8 - 2q_x] =120 $. The only way to realize it, is to assign values $4$ and $2$ to three vertices and value zero to the remaining vertices. Thus, we establish 
$ S^{(2)}_c \leq  \frac{120}{126} = \frac{20}{21}$. To show that the bound is tight, we provide a strategy in Fig. (\ref{fig4}) using four levels that satisfies clique-oblivious conditions. 
\end{proof}

\subsection*{YO-13 set}
\begin{figure}[H]
\centering
\includegraphics[scale=0.4]{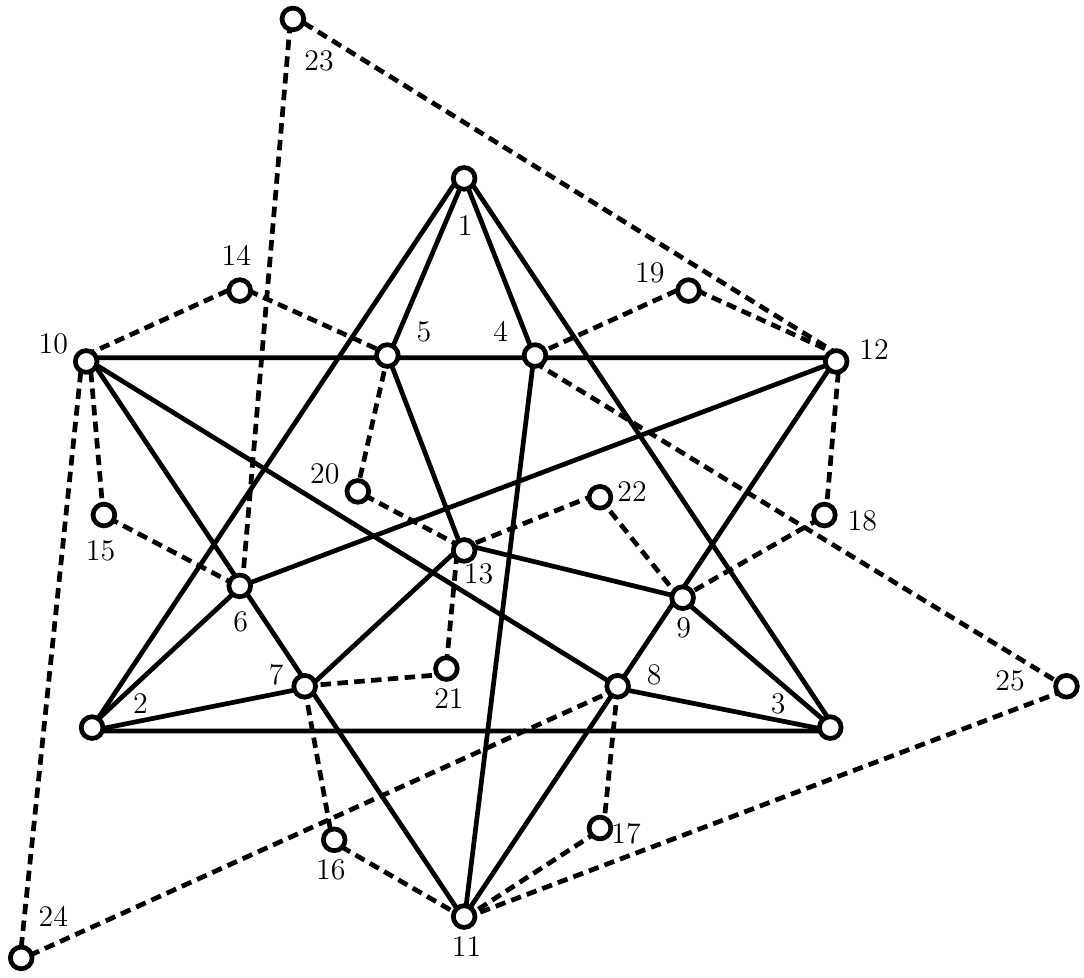}
\caption{Extended YO-13 graph with additional 12 vertices such that each vertex belongs to at least one 3-clique. The optimal encoding strategy is described in Table \ref{tab1}.}
\label{fig5}
\end{figure}

First we have to consider the extended YO-13 graph as shown in Fig. (\ref{fig5}). In the extended YO-13 graph, there are 25 vertices and 16 maximum cliques which correspond to oblivious variable $w\in \{1,\dots,16\}$,
\beq \label{13c}
&\Omega_1 =\{1,2,3\}, \Omega_2 =\{1,4,5\}, \Omega_3 =\{2,6,7\}, \nonumber \\ 
& \Omega_4 =\{3,8,9\}, \Omega_5 =\{5,10,14\}, \Omega_6 =\{6,10,15\}, \nonumber \\ 
& \Omega_7 =\{7,11,16\},  \Omega_8 =\{8,11,17\},  \Omega_9 =\{9,12,18\}, \nonumber \\ 
& \Omega_{10} =\{4,12,19\},  \Omega_{11} =\{5,13,20\}, \Omega_{12} =\{7,13,21\}, \nonumber \\
& \Omega_{13} =\{9,13,22\}, \Omega_{14} =\{6,12,23\}, \nonumber \\ 
&\Omega_{15} =\{8,10,24\}, \Omega_{16} =\{4,11,25\}. 
\eeq
 Since it is not a KS set, apart from clique-oblivious conditions, additionally we impose two more constraints. These two constraints are given by two subsets,
\be \label{13c1} \Omega_{17} = \{4,5,6,7,8,9\}, \ \Omega_{18}=\{10,11,12,13\}.\ee Notice that, since the following property is satisfied by the YO-13 set of vectors,
\be 
\frac{1}{6} \sum\limits_{i=4,5,6,7,8,9} |i\rangle \langle i| = \frac{1}{4} \sum\limits_{i=10,11,12,13} |i\rangle \langle i| = \frac{\mathbbm{1}}{3}, \ee 
the Quantum SIC strategy satisfies all the oblivious conditions and achieves the winning condition perfectly, that is, $S^{(2)}$ in \eqref{soc} is one.



\begin{prop}
The optimal classical value $S^{(2)}_c$ is 0.92 for the oblivious vertex equality problem based on extended YO-13 set.
\end{prop}
\begin{proof}
We follow similar method demonstrated in \textit{Proposition} \ref{prop:18}. First, we simplify the expression in \eqref{simexp} for the extended YO-13 set,
\beq
&S^{(2)}_c & = \frac{1}{58} \sum_{y} \max \left(c_y q_y, \sum_{ x\in N_y} c_x q_x \right) , \\
&& \text{where } \ c_x =
  \begin{cases}
    2/5, & x \in \{1,2,3\}\\
    4/7, & x \in \{4,...,13\} \\
    1/3, & x \in \{14,...,25\}.
  \end{cases}
\eeq 
Replacing the oblivious conditions given in \eqref{13c}-\eqref{13c1} on \eqref{cq}, we can infer that $q_x$ satisfy the following constraints,
\beq
&\forall x, q_x \geq 0; ~ q_1+q_2+q_3=q_1+q_4+q_5   =q_2+q_6+q_7  \nonumber \\ \nonumber
& =q_{3}+q_{8}+q_{9} =q_5+q_{10}+q_{14} =q_{6}+q_{10}+q_{15} \\ \nonumber
& = q_{7}+q_{11}+q_{16} = q_{8}+q_{11}+q_{17}  =q_{9}+q_{12}+q_{18} \\ \nonumber
& =q_{4}+q_{12}+q_{19} =q_{5}+q_{13}+q_{20} =q_{7}+q_{13}+q_{21} \\ \nonumber
& = q_{9}+q_{13}+q_{22} = q_{6}+q_{12}+q_{23}  = q_{8}+q_{10}+q_{24} \\ \nonumber
& =q_{4}+q_{11}+q_{25} =\frac{1}{2} (q_4+q_5+q_6+q_7+q_8+q_9) \\
& = \frac{3}{4} (q_{10} +  q_{11} + q_{12} + q_{13})= 3.
\eeq
Subsequently, one gets the extremal points of the polytope of $q_x$ by simple linear programing. In this case, there are 770 extremal points. The maximum value of $ \sum_{x} \max [q_x, \sum_{ y\in N_x} c_y q_y]= 53.4$ is obtained by considering all these extremal point. This leads to $S^{(2)}_c \lessapprox 0.92$. Finally we provide an encoding in table (\ref{tab1}) (see Appendix) using twelve levels, each of which satisfies the oblivious conditions, to show this bound is tight. 
\end{proof}

\subsection{Robustness in respect to white noise}
We discuss the robustness of the quantum advantage under the presence of white noise in both the aforementioned scenarios. 
Precisely, when Alice wants to send a quantum state $|\psi\rangle$, the actual communicated quantum system is $\nu |\psi\rangle \langle \psi| + (1-\nu) \mathbbm{1}/d$, where $\nu \in [0,1]$ is the white noise parameter. Note that, whenever a quantum communication protocol satisfies the oblivious constraints \eqref{Moc}, the noisy version also satisfies. Now, for the Quantum SIC strategy the observed probabilities, $p(0|x,y) = \nu + (1-\nu)/d$ for $x=y$, and $p(1|x,y)=\nu + (1-\nu)(d-1)/d$ for $y\in N_x$. A simple calculation leads to the modified quantum values of $S$ \eqref{sbd}-\eqref{soc} in two different communication scenarios as follows, 
\beq
S^{(1)} = \nu + \frac{(1-\nu)((d-1) \sum_x |N_x| + |G|)}{Nd}, \nonumber \\
S^{(2)} = \nu + (1-\nu)\left(\frac{d-1}{d} - \frac{\sum_x c_x(d-2)}{Nd}\right) .
\eeq 
Thus, for the persistence of quantum advantage these values should be greater than $S_c$. Subsequently, the following condition must hold for quantum advantage:\\
$(i)$ In communication with bounded dimension,
\be
\nu > 1 - \frac{(1-S^{(1)}_c)Nd}{N+(d-2)|G|}, \ \text{where } N = \sum_x |N_x| + |G|;
\ee 
$(ii)$ In oblivious communication,
\be
\nu > 1 - \frac{(1-S^{(2)}_c)Nd}{N + (d-2)\sum_x c_x}, \ \text{where } N = \sum_w |\Omega_w|.
\ee
In the former scenario with bounded dimension, the threshold visibility for both CEG-18 and YO-13 sets are approximately 0.92. While, in the OC, the threshold visibility for CEG-18 and extended YO-13 sets are 0.85 and 0.80 respectively. 

\subsection{Semi-device independent information processing using SIC sets}

While the fully device independent scenario rely on the quantum nonlocality involving one preparation and at least two measurement devices \cite{bellreview}, a notion of semi-device independent information processing is proposed in \cite{qkd}. This scenario constitutes a preparation device and a measurement device exactly same as the communication scenario presented in Section \ref{sec:ctbd}. It is characterized by the following two features:
\begin{enumerate}
 \item There is no assumptions on the internal features of the preparation and measurement devices, except the fact that the dimension of the physical systems (classical or quantum) produced by the preparation device is restricted to certain number. 
\item The choice of the inputs $x,y$ is independent of classical randomness shared between the two devices.  
\end{enumerate}
If the observed value of $S$ is greater than $S^{(1)}_c$ \eqref{ints}, then we conclude that these devices are genuinely quantum. Thus, as a direct consequence of \textit{Proposition} \ref{prop:sbd}, we know any SIC set of vectors can be used to demonstrate non-classicality in semi-device independent approach, and further possibly useful for quantum information processing protocols, for example, quantum key distribution \cite{qkd}, random number certification \cite{rc}. While, here we identify the applicability of SIC sets, the detailed quantitative analysis under the presence of eavesdropper are left for future work.

\section{Advantage in oblivious communication over classical channel implies preparation contextuality}\label{sec:pc} 
We point out that the classical bound for an OC task is also the optimal bound in all theories that satisfy the notion of preparation noncontextuality \cite{Spekkens2005}.
This generalizes the result obtained for parity oblivious multiplexing \cite{Spekkens09,banik,AC.etal,AA.etal,AH.etal,pan}. As we have shown KS sets yield advantage in OC, one may infer the consistency with previous result \cite{Spekkens2005} that KS contextuality for sharp measurements implies preparation contextuality. However, the presented approach connects KS sets to preparation contextuality via an operational task.

Consider the ontological model $ \{\Lambda, \mu, \xi\}$ of an operational theory, where $\Lambda$ is the ontic state space and $\mu(\lambda|P) \in [0,1]$ is the probability distributions over the ontic space for some preparation $P$. Given an ontic state $\lambda$, $\xi(Z|M,\lambda)$ denotes the probability for an outcome $Z$ of measurement $M$. Let $p(Z|P,M)$ be the observed probability of getting the outcome $Z$ of a measurement $M$ on a preparation $P$, then
$p(Z|M,P) = \sum_{\lambda\in\Lambda} \mu(\lambda|P) \xi(Z|M,\lambda).$ Denoting the set of all measurement outcome of a measurement $M$ by $\mathcal{Z}_M$,  and $\mathcal{M}$ as the set of all possible measurements, two different preparations $P$ and $P'$ are operationally equivalent if
\be
\forall Z\in\mathcal{Z}_M, \forall M\in \mathcal{M},\ p(Z|M,P) = p(Z|M,P').
\ee
In a preparation noncontextual ontological model, the ontic descriptions of two operationally equivalent preparations $P,P'$ are the same,  i.e.,
\begin{equation}\label{pnc}
\forall \lambda\in\Lambda,\ \mu(\lambda|P) = \mu(\lambda|P').
\end{equation} 
We also assume that the ontic probability distribution of a preparation which corresponds to a convex combination of two preparations, preserves the convexity at the ontic level.

In OC, the preparation device, which is possessed by Alice, performs a preparation $P_x$ for input $x$ and transmits to Bob. It is also relevant to define  the effective preparation corresponding to the oblivious variable $w \in [n_w]$, denoted by $P_{w}$, as the convex combination of all possible preparation $P_x$ with the probability distribution $p(x|w)$. The oblivious constraints imply that the statistics of all possible measurements for different $P_{w}$ remain the same:
\beq \label{oct}
& \forall Z \in \mathcal{Z}_M, \forall M \in \mathcal{M},\forall w,w' \in [n_w], \nonumber \\
& p(Z|P_{w},M) = p(Z|P_{w'},M). \eeq
In other words, the effective preparation $P_{w}$ for different $w$ are operationally equivalent. Thus, a preparation noncontextual theory implies,
\beq
& \forall \lambda,  \forall w,w'  \in [n_w], \quad  p(\lambda|P_{w}) = p(\lambda|P_{w'}).
\eeq 
Using the Bayes' rule, 
$ p(\lambda|P_w) = p(P_w|\lambda)p(\lambda)/ p(P_w)$ where $p(\lambda) = \sum_w p(\lambda|P_w) p(P_w)$, we further obtain,
\be \label{c5}
\forall w,w'  \in [n_w], \quad \frac{p(P_{w}|\lambda)}{p(P_w)} = \frac{p(P_{w'}|\lambda)}{p(P_{w'})}.
\ee
Let's say $p(P_{w}|\lambda)=k p(P_w)$, and by taking summation over $w$ on both side leads to $k=1$. Consequently, we have
\be \forall w, \ p(P_w|\lambda) = p(P_w).\ee
This means, even if the ontic state $\lambda$ is determined, it does not contain any information about $w$. \\
In a communication task, we seek to maximize some function of the observed probabilities $p(Z|P,M)=\sum_{\lambda} \mu(\lambda|P) \xi(Z|M,\lambda)$. Therefor it is sufficient to consider an ontic model with the existence of a measurement that determines $\lambda$.  Assuming that the space $\Lambda$ is countable, $p(Z|P,M)$ is further obtained by applying a stochastic map acting from $\Lambda$ to $\mathcal{Z}_M$. Thus, any observed probabilities of an ontological model can be reproduced by evaluating the ontic state $\lambda$ followed by some classical post processing.  

We arrive at two observations about a preparation noncontextual model in OC tasks: $(i)$ it is sufficient to consider ontic model with a measurement that determines the ontic state $\lambda$ followed by classical post-processing, $(ii)$ even if the ontic state $\lambda$ is determined, it does not contain any information about $w$. This is equivalent to the scenario where the preparation device communicates arbitrary number of distinct levels and each level is oblivious with respect to the oblivious variable, exactly as considered in the OC problem with classical channel. Thus, the classical bound in OC is also the bound for preparation noncontextual models, to wit, advantage in OC over classical channel signifies preparation contextuality.

\section{Conclusion}
The implication of this work is manifolded. First, the quantum advantage in communication: \textit{Equality problem} has been one of the central foci in communication complexity along with its applications in very-large-scale integration circuits, query complexity, data structure, and many more \cite{ccbook}. We show that pertaining to every logical proof of Kochen-Specker contextuality involving rank-one projectors there exists an equality problem with quantum advantage over classical channel in two distinct scenarios.  Due to the generality of our approach, this inference applies to every existing KS set of vectors as well as to those yet to be discovered. \\
Second, the operational significance of quantum contextuality: Test of single system contextuality is demanding to realize operationally. In particular, the compatibility loophole \cite{Guhne10} and the assumption of determinism \cite{Spekkens2014,KS2015,Xu.etal} are critically addressed. In regard to the advantageous quantum strategy, remarkably, the transmitted quantum states by the sender and measured quantum observables by the receiver simply correspond to the vectors that constitute the SIC set. 
This, in turn, manifests SIC as a sufficient resource in one-way communication tasks. 
Apart from the advantages in one-way communication, our results uncover the possibilities to implement semi-device independent information processing based on SIC sets. Moreover, we show preparation contextuality is the necessary resource for advantage in OC tasks. Taking into account the recent works \cite{amaral,saha}, one may attempt to provide a complete resource theory of quantum contextuality in the context of communication tasks.\\
Third, the quantum advantage in oblivious communication: By proposing the broad framework of OC, we provide a generalization to the previously studied parity oblivious multiplexing tasks \cite{Spekkens09,AC.etal,AA.etal,AH.etal,pan}, and a methodology to obtain the optimal strategy in classical communication. Unlike the communication scenario with bounded dimension, the quantum advantage in OC is unconditional on the amount of classical resource. As a consequence, it reveals a compelling notion of non-classicality in prepare-and-measure scenario for single quantum systems of finite dimension. 

One may consider the vertex equality problem imposing both the constraints of bounded dimension and oblivious information. Interestingly, quantum success will remain certain while the optimal classical value will be less or equal to the minimum of these two scenarios. 
Another task would be to compare the complexity for implementing classical and quantum strategies. 
In future, it will be worthwhile to consider other SIC sets and look for the optimal separation between quantum and classical success probability in both the scenarios.\\ 

\textit{Acknowledgments.---} 
We thank all the anonymous referees for helpful suggestions throughout the evaluation of the manuscript. DS thanks Anubhav Chaturvedi for comments. 
This work is supported by NCN grants 2016/23/N/ST2/02817, 2014/14/E/ST2/00020, John Templeton Foundation and FNP project First TEAM/2017-4/31.


\subsection*{Appendix}\label{app}

\begin{table}[H]
\centering
\bgroup
\def\arraystretch{1.35}
\begin{tabular}{c|c|c|c|c|c|c|c|c|c|c|c|c}
  & $A$ & $B$ &$C$ &$D$ &$E$ &$F$ &$G$ &$H$ &$I$ &$J$ &$K$ &$L$ \\
\hline
1 & $\frac{7}{36}$ & $\frac{7}{36}$ & $\frac{7}{36}$ & $\frac{7}{36}$ & $\frac{1}{36}$ & $\frac{1}{36}$ & $\frac{1}{36}$ & $\frac{1}{36}$ & $\frac{1}{36}$ & $\frac{1}{36}$ & $\frac{1}{36}$ & $\frac{1}{36}$ \\
2 & $\frac{1}{36}$ & $\frac{1}{36}$ & $\frac{1}{36}$ & $\frac{1}{36}$ & $\frac{7}{36}$ &  $\frac{7}{36}$ & $\frac{7}{36}$ & $\frac{7}{36}$ & $\frac{1}{36}$ & $\frac{1}{36}$ & $\frac{1}{36}$ & $\frac{1}{36}$\\
3 & $\frac{1}{36}$ & $\frac{1}{36}$ & $\frac{1}{36}$ & $\frac{1}{36}$ & $\frac{1}{36}$ &  $\frac{1}{36}$ & $\frac{1}{36}$ & $\frac{1}{36}$ & $\frac{7}{36}$ & $\frac{7}{36}$ & $\frac{7}{36}$ & $\frac{7}{36}$\\
4 & $\frac{2}{36}$ & 0 & 0 & $\frac{2}{36}$ & 0 & 0 & $\frac{8}{36}$ & $\frac{8}{36}$ &  0 & $\frac{8}{36}$ & 0 & $\frac{8}{36}$\\
5 & 0 & $\frac{2}{36}$ & $\frac{2}{36}$ & 0 & $\frac{8}{36}$ & $\frac{8}{36}$ & 0 & 0 & $\frac{8}{36}$ & 0 & $\frac{8}{36}$ & 0\\
6 & 0 & $\frac{8}{36}$ & 0 & $\frac{8}{36}$ & $\frac{2}{36}$ & 0 & 0 & $\frac{2}{36}$ & 0 & 0 & $\frac{8}{36}$ & $\frac{8}{36}$\\
7 & $\frac{8}{36}$ & 0 & $\frac{8}{36}$ & 0 & 0 & $\frac{2}{36}$ & $\frac{2}{36}$ & 0 & $\frac{8}{36}$ & $\frac{8}{36}$ & 0 & 0\\
8 & 0 & 0 & $\frac{8}{36}$ & $\frac{8}{36}$ & 0 & $\frac{8}{36}$ & 0 & $\frac{8}{36}$ & $\frac{2}{36}$ & 0 & 0 & $\frac{2}{36}$\\
9 & $\frac{8}{36}$ & $\frac{8}{36}$ & 0 & 0 & $\frac{8}{36}$ & 0 & $\frac{8}{36}$ & 0 & 0 & $\frac{2}{36}$ & $\frac{2}{36}$ & 0\\
10 & $\frac{1}{4}$ & $\frac{1}{36}$ & $\frac{1}{36}$ & $\frac{1}{36}$ & $\frac{1}{36}$ & $\frac{1}{36}$ & $\frac{1}{4}$ & $\frac{1}{36}$ & $\frac{1}{36}$ & $\frac{1}{4}$ & $\frac{1}{36}$ & $\frac{1}{36}$\\
11 & $\frac{1}{36}$ & $\frac{1}{4}$ & $\frac{1}{36}$ & $\frac{1}{36}$ & $\frac{1}{4}$ & $\frac{1}{36}$& $\frac{1}{36}$ & $\frac{1}{36}$ & $\frac{1}{36}$ & $\frac{1}{36}$ & $\frac{1}{4}$ & $\frac{1}{36}$\\
12 & $\frac{1}{36}$ & $\frac{1}{36}$ & $\frac{1}{4}$ & $\frac{1}{36}$ & $\frac{1}{36}$ & $\frac{1}{4}$ & $\frac{1}{36}$ & $\frac{1}{36}$ & $\frac{1}{4}$ & $\frac{1}{36}$ &  $\frac{1}{36}$& $\frac{1}{36}$\\
13 & $\frac{1}{36}$ & $\frac{1}{36}$ & $\frac{1}{36}$ & $\frac{1}{4}$ & $\frac{1}{36}$ & $\frac{1}{36}$& $\frac{1}{36}$ & $\frac{1}{4}$ & $\frac{1}{36}$ & $\frac{1}{36}$ & $\frac{1}{36}$ & $\frac{1}{4}$\\
14 & 0 & $\frac{6}{36}$ & $\frac{6}{36}$ & $\frac{8}{36}$ & 0 & 0 & 0 & $\frac{8}{36}$ & 0 & 0 & 0 & $\frac{8}{36}$\\
15 & 0 & 0 & $\frac{8}{36}$ & 0 & $\frac{6}{36}$ & $\frac{8}{36}$ & 0 & $\frac{6}{36}$ & $\frac{8}{36}$ & 0 & 0 & 0\\
16 & 0 & 0 & 0 & $\frac{8}{36}$ & 0 & $\frac{6}{36}$ & $\frac{6}{36}$ & $\frac{8}{36}$ & 0 & 0 & 0 & $\frac{8}{36}$\\
17 & $\frac{8}{36}$ & 0 & 0 & 0 & 0 & 0 & $\frac{8}{36}$ & 0 & $\frac{6}{36}$ & $\frac{8}{36}$ & 0 & $\frac{6}{36}$\\
18 & 0 & 0 & 0 & $\frac{8}{36}$ & 0 & 0 & 0 & $\frac{8}{36}$ & 0 & $\frac{6}{36}$ & $\frac{6}{36}$ & $\frac{8}{36}$\\
19 & $\frac{6}{36}$ & $\frac{8}{36}$ & 0 & $\frac{6}{36}$ & $\frac{8}{36}$ & 0 & 0 & 0 & 0 & 0 & $\frac{8}{36}$ & 0\\
20 & $\frac{8}{36}$ & $\frac{6}{36}$ & $\frac{6}{36}$ & 0 & 0 & 0 & $\frac{8}{36}$ & 0 & 0 & $\frac{8}{36}$ & 0 & 0\\
21 & 0 & $\frac{8}{36}$ & 0 & 0 & $\frac{8}{36}$ & $\frac{6}{36}$ & $\frac{6}{36}$ & 0 & 0 & 0 & $\frac{8}{36}$ & 0\\
22 & 0 & 0 & $\frac{8}{36}$ & 0 & 0 & $\frac{8}{36}$ & 0 & 0 & $\frac{8}{36}$ & $\frac{6}{36}$ & $\frac{6}{36}$ & 0\\
23 & $\frac{8}{36}$ & 0 & 0 & 0 & $\frac{6}{36}$ & 0 & $\frac{8}{36}$ & $\frac{6}{36}$ & 0 & $\frac{8}{36}$ & 0 & 0\\
24 & 0 & $\frac{8}{36}$ & 0 & 0 & $\frac{8}{36}$ & 0 & 0 & 0 & $\frac{6}{36}$ & 0 & $\frac{8}{36}$ & $\frac{6}{36}$\\
25 & $\frac{6}{36}$ & 0 & $\frac{8}{36}$ & $\frac{6}{36}$ & 0 & $\frac{8}{36}$ & 0 & 0 & $\frac{8}{36}$ & 0 & 0 & 0
\end{tabular}
\egroup
\caption{The probabilities of sending twelve levels $\{A,B,...,L\}$ for 25 different inputs are listed. The effective classical state is an equal mixture of twelve different levels. Considering the decoding strategy provided in \eqref{dsmain}, one can obtain $S^{(2)}_c \approx 0.92$.}
\label{tab1}
\end{table}



\begin{thebibliography}{99}

 \bibitem{Specker60}
 E. P. Specker,
 Dialectica \textbf{14}, 239 (1960).

\bibitem{Bell66}
 J. S. Bell,
 Rev. Mod. Phys. \textbf{38}, 447 (1966).

\bibitem{KS67}
 S. Kochen and E. P. Specker,
 J. Math. Mech. \textbf{17}, 59 (1967).

\bibitem{Cabello08}
 A. Cabello, \href{http://link.aps.org/doi/10.1103/PhysRevLett.101.210401}
{Phys. Rev. Lett. \textbf{101}, 210401 (2008).}

\bibitem{Cabello09}  P. Badziag, I. Bengtsson, A. Cabello, and I. Pitowsky, \href{https://link.aps.org/doi/10.1103/PhysRevLett.103.050401}
{Phys. Rev. Lett. \textbf{103}, 050401 (2009).}

\bibitem{YO12}
S. Yu and C. H. Oh, \href{http://journals.aps.org/prl/abstract/10.1103/PhysRevLett.108.030402}{Phys. Rev. Lett. \textbf{108}, 030402 (2012).}




\bibitem{CSW}
A. Cabello, S. Severini, and A. Winter, \href{http://link.aps.org/doi/10.1103/PhysRevLett.112.040401}
{Phys. Rev. Lett. \textbf{112}, 040401 (2014).}


\bibitem{ana} A. Acin, T. Fritz, A. Leverrier, and A. B. Sainz, \href{https://link.springer.com/article/10.1007/s00220-014-2260-1}{Comm. Math. Phys. {\bf 334}, 533 (2015).}

\bibitem{Spekkens2005}
R. W. Spekkens, \href{http://journals.aps.org/pra/abstract/10.1103/PhysRevA.71.052108}{Phys. Rev. A \textbf{71}, 052108 (2005).}


\bibitem{HWVE} M. Howard, J. J. Wallman, V. Veitch, and J. Emerson, \href{https://www.nature.com/nature/journal/v510/n7505/full/nature13460.html} {Nature \textbf{510}, 351 (2014).}

\bibitem{anders} J. Anders, and D. E. Browne, \href{https://link.aps.org/doi/10.1103/PhysRevLett.110.120401}{Phys. Rev. Lett. \textbf{102}, 050502 (2009). }


\bibitem{Raussendorf} R. Raussendorf, \href{https://link.aps.org/doi/10.1103/PhysRevA.88.022322}
{Phys. Rev. A \textbf{88}, 022322 (2013).}

\bibitem{Spekkens09} R. W. Spekkens, D. H. Buzacott, A. J. Keehn, B. Toner,
and G. J. Pryde, \href{https://link.aps.org/doi/10.1103/PhysRevLett.102.010401} {Phys. Rev. Lett. {\bf 102}, 010401 (2009).}


\bibitem{KGPLC} M. Kleinmann, O. Guhne, J. R. Portillo, J.-A. Larsson,
and A. Cabello, \href{http://stacks.iop.org/1367-2630/13/i=11/a=113011}
{New J. Phys. \textbf{13}, 113011 (2011).}

\bibitem{Cabello2018} A. Cabello, M. Gu, O. Guhne, and Z.-P. Xu, \href{https://link.aps.org/doi/10.1103/PhysRevLett.120.130401}{Phys. Rev. Lett. {\bf 120}, 130401 (2018).}

\bibitem{OG.etal}O. Guhne, C. Budroni, A. Cabello, M. Kleinmann, and J.-A. Larsson, \href{https://link.aps.org/doi/10.1103/PhysRevA.89.062107}
{Phys. Rev. A \textbf{89}, 062107 (2014). }

\bibitem{AG.etal} A. Grudka, K. Horodecki, M. Horodecki, P. Horodecki, R. Horodecki, P. Joshi, W. Klobus, and A. Wojcik,
\href{https://link.aps.org/doi/10.1103/PhysRevLett.112.120401}
{Phys. Rev. Lett. \textbf{112}, 120401 (2014).}

\bibitem{arvind} J. Singh, K. Bharti, and Arvind, \href{https://link.aps.org/doi/10.1103/PhysRevA.95.062333}{Phys. Rev. A \textbf{95}, 062333 (2017).}

\bibitem{schmid} D. Schmid, and R. W. Spekkens, \href{https://link.aps.org/doi/10.1103/PhysRevX.8.011015} {Phys. Rev. X {\bf 8}, 011015 (2018).}


\bibitem{Brassard05}
G. Brassard, A. Broadbent, and A. Tapp, Alain,
\href{http://dx.doi.org/10.1007/s10701-005-7353-4} {Foundations of Physics, \textbf{35}, 1877 (2005).}

\bibitem{PC.etal} P. J. Cameron, A. Montanaro, M. W. Newman, S. Severini, and A. Winter, \href{https://www.emis.de/journals/EJC/Volume_14/PDF/v14i1r81.pdf}
{Electronic Journal of Combinatorics \textbf{14}, 1 (2007).}

\bibitem{Abramsky} S. Abramsky, R. S. Barbosa, N. Silva, O. Zapata, \href{https://arxiv.org/abs/1705.07310} {arXiv: 1705.07310 [cs.LO].}

\bibitem{CLMW} T. S. Cubitt, D. Leung, W. Matthews, and A. Winter, \href{https://link.aps.org/doi/10.1103/PhysRevLett.104.230503} {Phys. Rev.
Lett. \textbf{104}, 230503 (2010).}

\bibitem{KCK} P. Kurzynski, A. Cabello, and D. Kaszlikowski, \href{https://link.aps.org/doi/10.1103/PhysRevLett.112.100401} {Phys. Rev.
Lett. \textbf{112}, 100401 (2014).}

\bibitem{SR} D. Saha, and R. Ramanathan, \href{https://link.aps.org/doi/10.1103/PhysRevA.95.030104} {Phys. Rev. A \textbf{95}, 030104 (R) (2017).}


\bibitem{Cabello99}
A. Cabello, J. M. Estebaranz, and G. Garcia-Alcaine, \href{http://www.sciencedirect.com/science/article/pii/037596019600134X}{Phys. Lett. A \textbf{212}, 183 (1996).}



\bibitem{qkd} M. Paw\l{}owski and N. Brunner, \href{https://link.aps.org/doi/10.1103/PhysRevA.84.010302} {Phys. Rev. A {\bf 84}, 010302(R) (2011).}

\bibitem{rc} H-W. Li, Z.-Q. Yin, Y.-C. Wu, X.-B. Zou, S. Wang, W. Chen, G.-C. Guo, and Z.-F. Han, \href  {https://link.aps.org/doi/10.1103/PhysRevA.84.034301}
 {Phys. Rev. A {\bf 84}, 034301 (2011).}




\bibitem{brassard} G. Brassard, 
\href{https://doi.org/10.1023/A:1026009100467}{Foundations of Physics  \textbf{33}, 1593 (2003).}

\bibitem{wolf} R. de Wolf,
\href{https://doi.org/10.1016/S0304-3975(02)00377-8}{Theoretical Computer Science \textbf{287}, 337 (2002).}

\bibitem{reviewCC} H. Buhrman, R. Cleve, S. Massar, R. de Wolf, \href{https://link.aps.org/doi/10.1103/RevModPhys.82.665} {Rev Mod Phys \textbf{82}, 665 (2010)}.

\bibitem{dw1} S. Wehner, M. Christandl, and A. C. Doherty, \href{https://link.aps.org/doi/10.1103/PhysRevA.78.062112}
 {Phys. Rev. A {\bf 78}, 062112 (2008).}

\bibitem{dw2} R. Gallego, N. Brunner, C. Hadley, and A. Acin, \href {https://link.aps.org/doi/10.1103/PhysRevLett.105.230501} {Phys. Rev. Lett. {\bf 105}, 230501 (2010).}






\bibitem{crypto} C. Crepeau. "Equivalence between two flavours of oblivious transfer". In Advances in Cryptology: CRYPTO '87, vol. 293 of Lecture Notes in Computer Science, pages 350–354. Springer, 1988.

\bibitem{qcrypto} C. H. Bennett, G. Brassard, C. Crepeau, and M.-H. Skubiszewska, \href{https://link.springer.com/chapter/10.1007/3-540-46766-1_29}{Advances in Cryptology-CRYPTO'91, \textbf{576}, 351 (1991).}


\bibitem{qcrypto1} C. Crepeau,  \href{http://dx.doi.org/10.1080/09500349414552291}{J. Mod. Opt. \textbf{41}, 2445 (1994).}

\bibitem{marcin} M. Plesch, M. Paw\l{}owski, and M. Pivoluska, \href{https://link.aps.org/doi/10.1103/PhysRevA.95.042324}{Phys. Rev. A \textbf{95}, 042324 (2017).}


\bibitem{ppc} Andrew Chi-Chih Yao, \href{doi.ieeecomputersociety.org/10.1109/SFCS.1986.25}{27th Annual Symposium on Foundations of Computer Science (FOCS), 162-167 (1986).}



\bibitem{banik} M. Banik, S. S. Bhattacharya, A. Mukherjee, A. Roy, A. Ambainis, and A. Rai, \href{https://link.aps.org/doi/10.1103/PhysRevA.92.030103}{Phys. Rev. A \textbf{92}, 030103 (R) (2015).}

\bibitem{AC.etal}
A. Chailloux, I. Kerenidis, S. Kundu, and J. Sikora, \href{http://stacks.iop.org/1367-2630/18/i=4/a=045003}{New J. Phys. \textbf{18}, 045003 (2016).}

\bibitem{AA.etal} A. Ambainis, M. Banik, A. Chaturvedi, D. Kravchenko, and A. Rai, \href{https://doi.org/10.1007/s11128-019-2228-3}{Quantum Inf Process \textbf{18}, 111  (2018)}.

\bibitem{AH.etal} A. Hameedi, A. Tavakoli, B. Marques, and M. Bourennane, \href {https://link.aps.org/doi/10.1103/PhysRevLett.119.220402} {Phys. Rev. Lett. \textbf{119}, 220402 (2017).}.

\bibitem{pan} S. Ghorai, and A. K. Pan, \href{https://link.aps.org/doi/10.1103/PhysRevA.98.032110}{Phys. Rev. A \textbf{98}, 032110 (2018).} 

\bibitem{saha} D. Saha, and A. Chaturvedi, \href{https://link.aps.org/doi/10.1103/PhysRevA.100.022108}{Phys. Rev. A \textbf{100}, 022108 (2019). }





\bibitem{Cabello11} A. Cabello, M. Kleinmann, and C. Budroni, \href{https://link.aps.org/doi/10.1103/PhysRevLett.114.250402} {Phys. Rev. Lett. \textbf{114}, 250402 (2015).}

\bibitem{RH} R. Ramanathan, and P. Horodecki, \href{https://link.aps.org/doi/10.1103/PhysRevLett.112.040404} {Phys. Rev. Lett. \textbf{112}, 040404 (2014).}

\bibitem{bellreview} N. Brunner, D. Cavalcanti, S. Pironio, V. Scarani, and S. Wehner, \href{https://link.aps.org/doi/10.1103/RevModPhys.86.419}{Rev. Mod. Phys. {\bf 86}, 419 (2014).}




%



\bibitem{ccbook} E. Kushilevitz, and N. Nisan, \textit{Communication Complexity}, Cambridge Univ Press,
Cambridge, UK (2006).

\bibitem{Guhne10} O. Guhne, M. Kleinmann, A. Cabello, J.-A. Larsson, G. Kirchmair, F. Zahringer, R. Gerritsma, and C. F. Roos, \href {https://link.aps.org/doi/10.1103/PhysRevA.81.022121} {Phys. Rev. A \textbf{81}, 022121 (2010).}

\bibitem{Spekkens2014}
R. W. Spekkens, \href{http://dx.doi.org/10.1007/s10701-014-9833-x}
{Found. Phys. \textbf{44}, 1125–1155 (2014).}

\bibitem{KS2015}
R. Kunjwal and R. W. Spekkens, \href{http://journals.aps.org/prl/abstract/10.1103/PhysRevLett.115.110403}{Phys. Rev. Lett. \textbf{115}, 110403 (2015).}

\bibitem{Xu.etal}
Z.-P. Xu, D. Saha, H.-Y. Su, M. Paw\l{}owski, and J.-L. Chen, \href{https://link.aps.org/doi/10.1103/PhysRevA.94.062103}{Phys. Rev. A \textbf{94}, 062103 (2016).}



\bibitem{amaral} C. Duarte, and   B. Amaral, \href{https://doi.org/10.1063/1.5018582}{J. Math. Phys. \textbf{59}, 062202 (2018).}





\end{thebibliography}
\end{document}